\documentclass{article}
\pdfoutput=1
\usepackage[utf8]{inputenc}
\usepackage[numbers]{natbib}
\usepackage{graphicx} 
\usepackage{url}
\usepackage{hyperref}
\usepackage{authblk}

\usepackage[margin=2cm]{geometry}

\title{Multi-Source Social Feedback of Online News Feeds}

\author{Nuno Moniz\thanks{Email: nmmoniz@inescporto.pt}}
\author{Luís Torgo\thanks{Email: ltorgo@dcc.fc.up.pt}}
\affil{LIAAD - INESC Tec\\ FCUP - DCC, University of Porto}

\begin{document}

\maketitle

\abstract{The profusion of user generated content caused by the rise of social media platforms has enabled a surge in research relating to fields such as information retrieval, recommender systems, data mining and machine learning. However, the lack of comprehensive baseline data sets to allow a thorough evaluative comparison has become an important issue. In this paper we present a large data set of news items from well-known aggregators such as Google News and Yahoo! News, and their respective social feedback on multiple platforms: Facebook, Google+ and LinkedIn. The data collected relates to a period of 8 months, between November 2015 and July 2016, accounting for about 100,000 news items on four different topics: economy, microsoft, obama and palestine. This data set is tailored for evaluative comparisons in predictive analytics tasks, although allowing for tasks in other research areas such as topic detection and tracking, sentiment analysis in short text, first story detection or news recommendation.\footnote{Link to dataset: \url{http://www.dcc.fc.up.pt/~nmoniz/MultiSourceNews}}.}


\section{Summary}\label{sec:summary}

Social media is a broad term encompassing forums, blogs, video sharing websites, collaborative coding platforms and social networking platforms (e.g. Facebook, Twitter)~\cite{Tsagkias2012}. Their common denominator is the possibility for users to generate and share content, i.e. web content. The web content generated by social media platforms, which is also referred to as social media data, is diverse, including text, images, audio and videos, and the impact of social media platforms is enormous when considering the volume of web traffic and content it generates. 

Consider the following statistics from three of the most popular platforms: \textit{i)} the number of users in Youtube is almost a third of all Internet users (over a billion users)~\cite{Youtube2016}, representing a significant amount of all Internet traffic~\cite{Gursun2011}; \textit{ii)} in September 2016, Facebook had an average of more than a billion daily active users~\cite{Facebook2016}; and Twitter generated over a billion visits to sites using embedded tweets~\cite{Twitter2016}.

Despite the profusion of social media data, the context in which it is generated and shared presents several factors that may facilitate or hinder the ability to use such data~\cite{Liu2016}. On one hand, social media data is massive and linked. On the other hand, it is commonly noisy, sparse, informal and biased. Additionally, it is heterogeneous, partial, asymmetrical and multi-sourced. This combination of factors illustrates well the difficulty of using web content and their impact in reducing the ability to successfully disclose its full potential.

The potential of social media data is immense. Its broad definition concerns a very diverse set of data types with particular properties. For example, online news are a very dynamic type of web content, due to the information it conveys and its impact on everyday life. As such, new events and stories are constantly being published as well as updates on old stories. This causes news items to have a very short life span~\cite{Morales2012,Yang2011}. In contrast, online videos have a much longer alive-time, enduring for weeks or months~\cite{Szabo2010}.

For over a decade, researchers have taken advantage of the profusion of user generated data in social media to tackle a diverse set of tasks concerning information retrieval, recommender systems, data mining and machine learning. Efforts by many data proprietaries have been made in order to allow access to this enormous amount of data. However, for several fields, the lack of ready-to-use, accessible, comprehensive, detailed and large data sets that would serve as basis for comparison between approaches, is a major issue. This is the main motivation for our contribution. 

In this paper we describe a large scale data set of news items from multiple official media sources and their respective social feedback in multiple social media platforms. Our objective is to establish this data set as a well-known baseline for research efforts pertaining to various fields of research concerning web content, and specifically predictive analytics.


\section{Data Description}\label{sec:datadesc}

News data can be obtained from two types of sources: \textit{i)} official media, and \textit{ii)} social media sources. The first type of source, official media, relates to the origin of news items and their content. It may also provide an indication of the items' relevance according to the official media source, denoted by its ranking position. The second, social media, is a medium used to measure the attention received by news items, i.e. popularity. In previous work, the provenance of the data used can be framed in one of three settings: \textit{i)} solely using official media sources (\textit{e.g.} \cite{Wu2007,Lerman2010b,Szabo2010}), \textit{ii)} solely using social media sources (\textit{e.g.} \cite{Schulman2016,Petrovic2013}), or \textit{iii)} using data from both official and social media sources (\textit{e.g.} \cite{Lerman2010a,Hong2011,Morales2012}).

Official media sources include legacy media\footnote{Common expression to denote traditional media outlets.} outlets (\textit{e.g.} The Washington Post), news aggregation platforms (\textit{e.g.} Digg, Slashdot) and news recommender systems (\textit{e.g.} Google News, Yahoo! News). Most of previous work solely concerning official sources was focused on the second, with emphasis on the Digg platform. Since one of our objectives is to allow the analysis of recommendation and/or learning to rank approaches, this data set is based on the third type of official sources: news recommender systems. This choice has the benefits of \textit{i)} providing a potentially large list of items, and of \textit{ii)} presenting a multitude of news outlets, including those that are not considered as being legacy media. 

Regarding social media sources of news data, research has shown that the micro-blogging platform Twitter covers almost all news-wire events but the opposite is not true~\cite{Petrovic2013}. Furthermore, Osborne and Dredze~\cite{Osborne2014} have shown that Twitter is the preferred medium for breaking news, almost consistently leading Facebook or Google+. However, this claim has since grown out of date, as it has been shown~\cite{Osborne2014} that Facebook is now the leading platform in accessing and sharing news, followed by Twitter. Newman et al.~\cite{Newman2016} provide a survey of over 50,000 people in 26 countries concerning news consumption. It shows that the most popular platform is Facebook, followed by Youtube, Twitter, Whatsapp, Google+, LinkedIn and Instagram. The data set presented in this paper does not include the sources Youtube, Whatsapp, Instagram and Twitter. Concerning the first three, for technical reasons, such platforms either do not allow the sharing of news items (although allowing it in comments) as in the case of Youtube and Instagram, or are essentially focused on direct messaging, significantly reducing the dynamics of popularity and social spread. In the case of Twitter, the decision was related to the deprecation of its API functionality allowing for the extraction of such data\footnote{Twitter API: \url{https://developer.twitter.com/en/docs}. The \textit{count} method was deprecated on 20$^{th}$ of November, 2015.}. 

Nonetheless, considering the wide use of Twitter data in the context social media data research, it should be noted that other alternatives would be possible, such as using the search functionality of Twitter API or obtaining archive access services. The first possibility was discarded due to the loss of information implied in such functionality, since Twitter only allows the access to 1\% of its data stream\footnote{Adjustments to Streaming API sample volumes (Twitter): \url{https://twittercommunity.com/t/potential-adjustments-to-streaming-api-sample-volumes/31628}}. The second possibility was also discarded due to the heavy financial implications regarding such archive access services.

In order to allow the analysis and discussion of the interplay between these two types of data sources, the data set presented in this paper uses data from both types of sources: official and social media. The information portrayed in such sources is used and interpreted differently. From official data sources descriptors of news items are extracted. As for social media sources, these are used to obtain the popularity of news, which may be denoted by different signals in each source.

\section{Data Set}\label{sec:dataset}

This data set is a multi-source data set, using two official media sources (Google News and Yahoo! News) and three social media sources (Facebook, Google+ and LinkedIn). It contains news-related data concerning four topics: \emph{economy, microsoft, obama,} and \emph{palestine}. These topics were chosen ad-hoc, based on two factors: their worldwide popularity and constant activity, and the fact that they report to different types of entities (sector, company, person, and country, respectively). Notwithstanding, this choice raises some caveats. Being limited to a small number of topics might undermine conclusions concerning the ability to generalize the predictive ability reported in experiments. On the other hand, it does provide a deep insight into topics that have a daily activity of high magnitude, in opposition to having a news sample of a variety of topics. By approaching the development of data sets as proposed, possible problems related to context-sensitivity (\textit{i.e.} topic) characteristics of text-based tasks~\cite{Blitzer2007} are also tackled. Furthermore, due to the technical restrictions, such as access limitations per hour that both Google News and Yahoo! News applies, extending the list of topics would not be possible unless the time intervals of access would be greater. In this light, we chose to maintain the previously mentioned list of topics.

Concerning the data horizon, previous work has differed on the active time that one should consider news items relevant, since their publication, although agreeing that such items have a very short lifetime in terms of capturing the attention of users, in comparison to other types of web content, i.e. video. For example, researchers have shown evidence for alternative values such as two days~\cite{Morales2012,Lakkaraju2011}, four days~\cite{Yang2011,Bandari2012} and 30 days~\cite{Tatar2014} concerning the alive-time for which news may receive the attention of users, leading to such items being shared on social media platforms. However, based on the analysis provided by these works, most of the popularity dynamics of news develop in the first day, although showing that in some cases it may develop for more time. Nonetheless, in the latter cases, the increase is very residual in terms of proportion. Given this, an active time for the news items of two days is established, since their publication. This means that the popularity evolution of each item is stored for this period, since their appearance in official media sources.

Finally, it should be noted that a stochastic view of the popularity concept is assumed (i.e. aggregate behaviour~\cite{Tatar2014}), considering all publications from every user equally. Different approaches have been proposed, such as those focused on determining the number of “retweets” (Twitter functionality to re-publish a tweet) a given tweet will obtain~\cite{Suh2010,Zaman2014,Hong2011} or those using data concerning the social network of individual users to predict the popularity of content they generate~\cite{Gupta2012}. The data composed in these data sets is not focused on the popularity of content generated by a single or a given group of users, but on the general popularity of content in social media platforms, allowing for a source-wide perception of news stories' popularity.

The data set is divided in two sets of data files. The first set concerns news data, and is represented by a single data file, containing 11 variables that identify and describe the news items. The second set of data files relate to social feedback data concerning the news items, and is composed by 12 data files with 145 variables each. These variables contain the information of the evolution of the popularity of news items for a period of two days. In the following sections, the two sets of data files are described, and the methods used to obtain the data are detailed in Section~\ref{sec:methods}.

\subsection{News Data}\label{subsec:newsdata}

In this section, the data file concerning the description of news items is detailed. The data descriptors are based on information obtained by querying the official media sources Google News and Yahoo! News. Each news item is described by 11 variables, which are explained in Table~\ref{tbl:newsfile}.

\begin{table}[h]
\begin{center}
\scriptsize
\caption{Name, type of data and description of data variables in news data file.}\label{tbl:newsfile}
\begin{tabular}{l l p{12cm}}
\textbf{Variable}                & \textbf{Type}    & \textbf{Description}                                                                                                                                                              \\
\hline
\textbf{IDLink}                    & \textit{numeric}  & Unique identifier of news items \\

\textbf{Title}                   & \textit{string}  & Title of the news item according to the official media sources \\

\textbf{Headline}                & \textit{string}  & Headline of the news item according to the official media sources \\

\textbf{Source}                   & \textit{string} & Original news outlet that published the news item \\

\textbf{Topic}                     & \textit{string} & Query topic used to obtain the items in the official media sources \\

\textbf{PublishDate}            & \textit{timestamp}  & Date and time of the news items' publication \\

\textbf{SentimentTitle}                & \textit{numeric}  & Sentiment score of the text in the news items' title \\

\textbf{SentimentHeadline}                & \textit{numeric}  & Sentiment score of the text in the news items' headline \\

\textbf{Facebook}                & \textit{numeric}  & Final value of the news items' popularity according to the social media source Facebook \\

\textbf{GooglePlus}                & \textit{numeric}  & Final value of the news items' popularity according to the social media source Google+ \\

\textbf{LinkedIn}                & \textit{numeric}  & Final value of the news items' popularity according to the social media source LinkedIn \\

\hline
\end{tabular}
\end{center}
\end{table}

In addition to the original information retrieved from the official media sources, we chose to include the additional information of the sentiment scores of both the title and headline of the news items. Sentiment analysis, also known as opinion mining, is a field of text mining which studies people's opinions, judgments and ideas towards entities~\cite{Liu2012}. In this data set, the process to obtain these sentiment scores was carried out by applying the framework of the \textbf{qdap} R package~\cite{Rinker2013} with default parametrization. It should be made clear that these results were obtained by an off-the-shelf approach, and as such, its values should not be interpreted as being verified as to its accuracy. The choice of adding such information was in order to enrich the data set and make clear the options for further research by data set users.

In previous work, several authors have used social network features, such as its structure \cite{Jamali2009, Tsur2012} the number of followers and/or followees~\cite{Lerman2008}. The reasons for not including such information in this data set are two. First, using social network features in a broad spectrum prediction-based system raises the issue of access to data. Although it is possible to relax this issue, by obtaining a sample of data which is used to derive and test prediction models, its extensive use would virtually require full access to real-time data. Unfortunately, it is highly unlikely that this option is available to anyone other than the data owners. Secondly, handling data on the connections that users establish with other users, as well as harnessing and using their interactions, raises issues related to privacy~\cite{Garcin2013}.

\begin{figure}[!h]
\centering
\includegraphics[width=15cm]{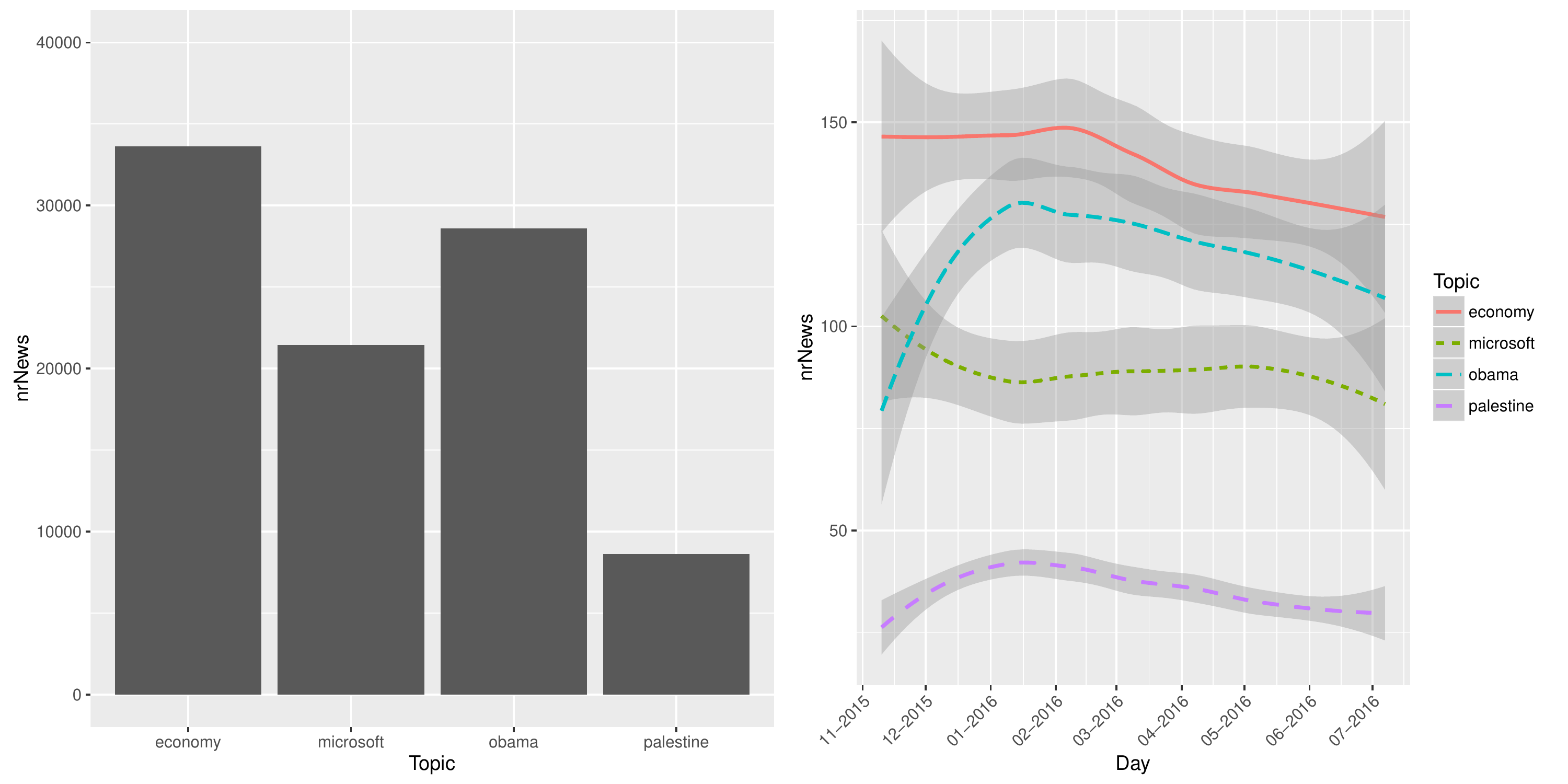}
\caption{Number of news from both Google News and Yahoo! News (left) and a smoothed approximation of the amount of news per day, for each topic.}\label{fig:newstopicday_multi}
\end{figure}

Figure~\ref{fig:newstopicday_multi} shows the global number of news per topic (left) and a smoothed approximation of the amount of news per day for each topic (right). The total amount of news retrieved in both official media sources is 93,239.

\subsection{Social Feedback Data}\label{subsec:feedbackdata}

The second set of data files concerns the evolution of news items' popularity in the various social media sources. This set is composed of 12 data files, for all combinations of the four topics (economy, microsoft, obama and palestine) and the available social media sources Facebook, Google+ and LinkedIn. Each case is described by 145 variables: a unique identifier, and 144 measurements of popularity in 20 minute intervals for a total of 2 days.

\begin{figure}[!h]
\centering
\includegraphics[width=15cm]{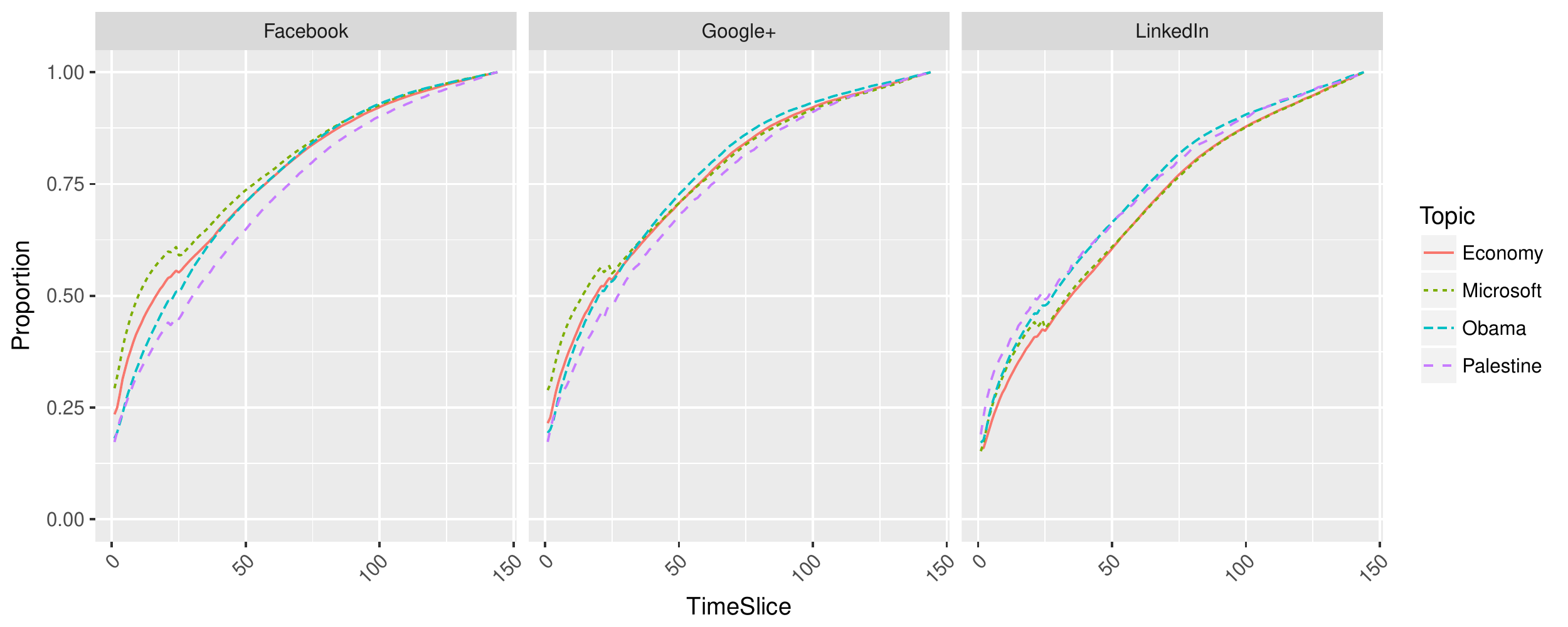}
\caption{Evolution of popularity (as proportion of final popularity) in each topic, for social media sources. Each time slice represents a 20 minute period.}\label{fig:q1}
\end{figure}

The evolution of popularity in each of the combination of social media source-topic available in the data sets is different. Figure~\ref{fig:q1} illustrates the evolution of popularity for each of the query periods in the data set, referred to as time slices. Results are depicted as the average proportion between the popularity scores at each given time slice, and the respective final popularity scores.

Based on the results depicted in Figure~\ref{fig:q1}, it is observed that in most social media sources, news items obtain close to half of their final popularity in a short amount of time: Facebook, Google+ and LinkedIn show that in the first moments after publication, news items quickly obtain on average close to 20\% ($22\%$, $21.8\%$ and $16.8\%$, respectively) of their final popularity.

\begin{figure}[!h]
\centering
\includegraphics[width=11cm]{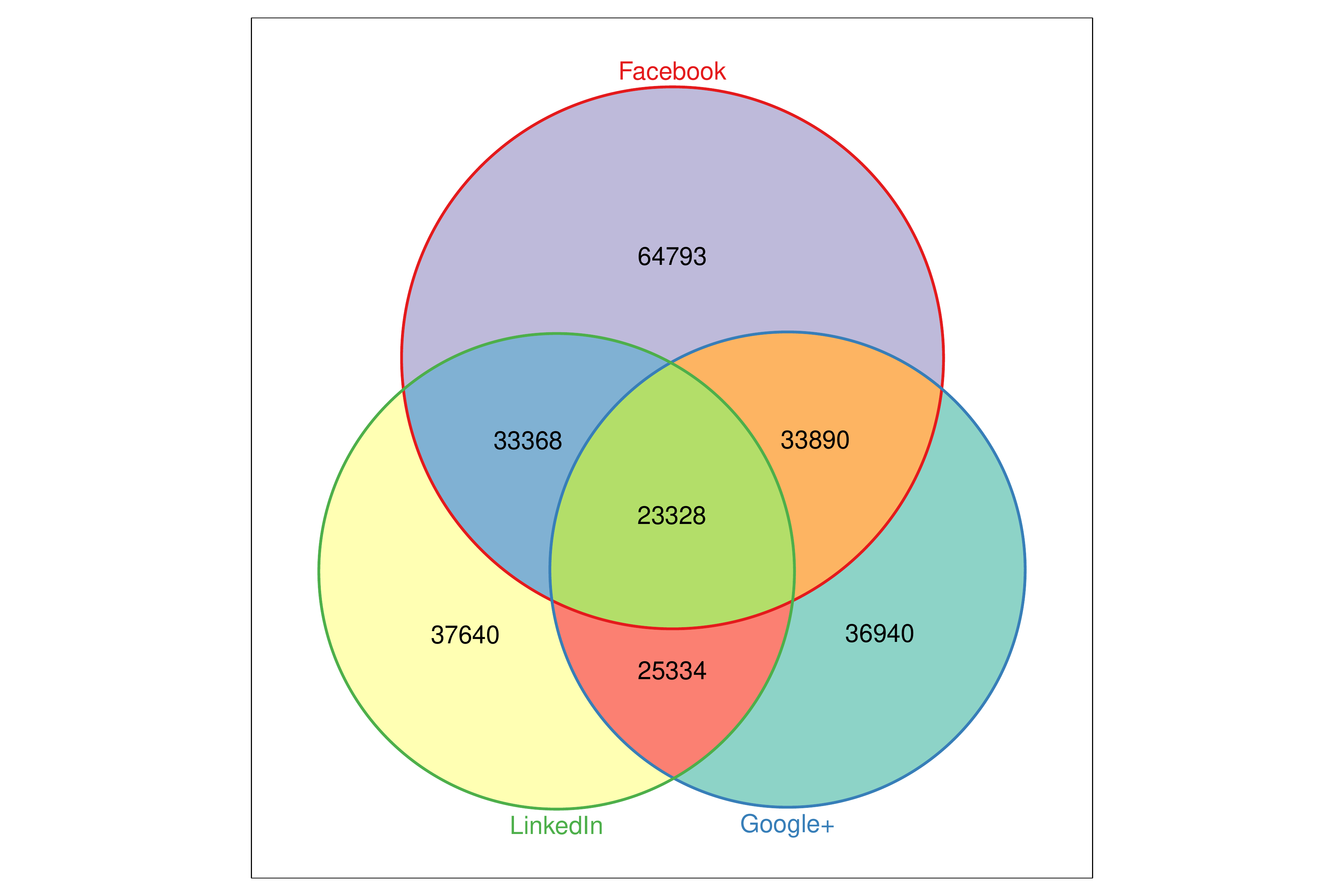}
\caption{Venn diagram of published news items in social media sources.}\label{fig:vennsocial}
\end{figure}

In Figure~\ref{fig:vennsocial} the intersection of the sets of news that were published in each of the social media sources is illustrated. Results show that most of the news published in Google+ or LinkedIn are also published in Facebook ($91.7\%$ and $88.7\%$ respectively). Conversely, data shows that roughly a third ($36\%$) of news published in Facebook are also published in both Google+ and LinkedIn. Additionally, only 2,006 ($7.9\%$) news stories were published in both Google+ and LinkedIn, and not on Facebook.

\section{Methods}\label{sec:methods}

The official media sources Google News and Yahoo! News were queried, during a period of approximately eight months (November 10$^{th}$, 2015 until July 7$^{th}$ 2016), for each of the four topics (\textit{economy}, \textit{microsoft}, \textit{obama} and \textit{palestine}). The queries were done simultaneously, in 20 minute intervals. For each query, the top-100 recommended news of the respective official media sources were collected. For each recommended news item, the following information was collected: title, headline, publication date and the news outlet. 

In order to deal with duplicated cases, the following procedure was applied. First, cases with the same title, headline and from the same news outlet, are grouped. Second, the oldest case is kept and the remaining are removed from the data set. Finally, the identifiers of the removed cases are replaced by that of the case that was kept. 

In addition, concerning the data quality from official media sources, we should note that some of the news items' title and headline may be truncated (represented by the use of ''...''), which is an option of the official media sources Google News and Yahoo! News. Alternatives to this shortcoming would include following the URL of each news and creating tailored components for each news outlet. Considering the magnitude of news outlets in our data set, we found this to be impractical. Also, standard operations for cleaning the textual variables were applied, namely the removal of double, trailing and leading white spaces.

After retrieving the query-data from the official sources, the popularity of all known news items, with an alive time below the defined period of two days, is obtained from the social media sources Facebook, Google+ and LinkedIn, simultaneously. Considering the differences between each social media source, the procedures for obtaining popularity are different. They are described as follows.

\begin{itemize}
\item For obtaining information from Facebook, the Facebook Graph API\footnote{The Graph API: \url{https://developers.facebook.com/docs/graph-api}} is used, by querying for information concerning the URL of each news item. The data retrieved reports the number of shares concerning each unique URL, which is used as a popularity measure. In 11,602 cases ($12.4\%$) it was not possible to obtain the number of shares, and in 26,919 cases ($28.9\%$) the news items' were not shared on Facebook;

\item The social media platform Google+ does not allow to obtain the number of shares of a given URL, due to technical restrictions. Nonetheless, it allows one to check the number of times users have ''liked'' the URL's. Despite the differences with other social media sources, it is nonetheless a valid metric of received attention by news stories. This process is carried out by querying a public end-point\footnote{The number of ''+1'' received by a given URL in Google+ is obtained by appending the respective URL to \url{https://plusone.google.com/_/+1/fastbutton?url=}. This approach is no longer available.} in order to obtain the amount of ''+1'' (similar to ''like'' in Facebook) a given URL received. In 5,744 cases ($6.2\%$) it was not possible to obtain the number of ''+1'', and in 55,114 cases ($59.1\%$) the news items' did not obtain any ''+1'';

\item Finally, concerning the LinkedIn platform, the number of times each news story URL was shared is obtained by querying its public end-point\footnote{The number of times a given URL was shared in LinkedIn is obtained by appending the respective URL to \url{https://www.linkedin.com/countserv/count/share?format=json&url=}.}, designed for such purposes. Concerning the overall statistics of the news items' presence in the platform, in 5,745 cases ($6.2\%$) it was not possible to obtain information concerning the news item URL. In addition, in 54,413 cases ($58.4\%$) the news were not shared on the LinkedIn platform.
\end{itemize}

In some cases, the value of the popularity of news items at a given moment (i.e. timeslice) was not acquirable. These cases are denoted with the value $-1$, which are mostly associated to scenarios where the items are suggested in official media sources after two days have passed since its publication in the original news outlet. Such situations represent $12.4\%$ of cases concerning the final popularity of items according to the social media source Facebook, and $6.2\%$ of cases for both Google+ and LinkedIn sources. 

\begin{figure}[!h]
\centering
\includegraphics[width=15cm]{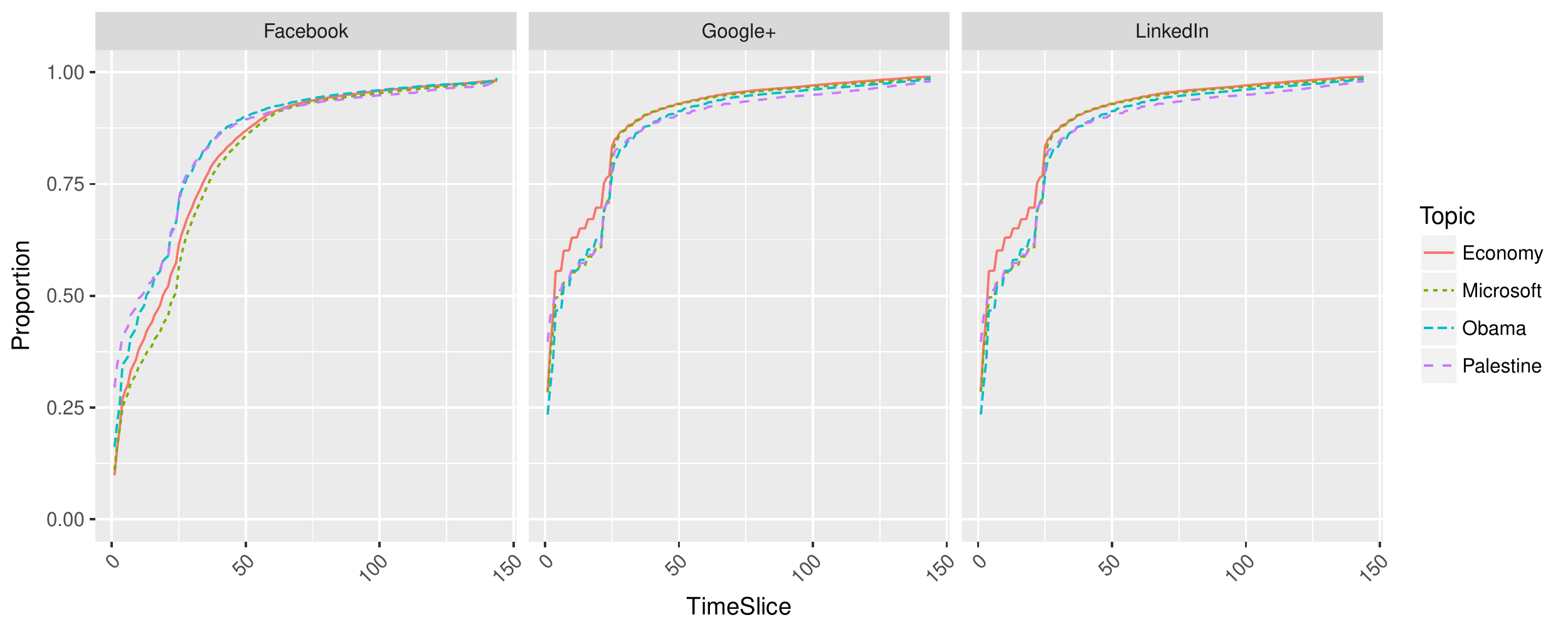}
\caption{Evolution of available information in each time slice for all topics and social media sources.}\label{fig:availableData}
\end{figure}

In addition, a common situation is when the news items appeared in the official media sources' recommendations with some delay w.r.t. their original publication time. This leads to some cases only having popularity data since a given time slice. Concerning this issue of the available data according to each time slice (i.e. query period of 20 minutes) of news items, Figure~\ref{fig:availableData} depicts the evolution of the available data proportion for all social media sources and query topics. It illustrates the proportion of cases for which popularity data is available in each time slice.

Finally, it is important to clarify the ethical component of this data set as it pertains to the use of publicly-available data generated by users in social media platforms~\cite{Ethics2012}. As such, we should stress that, as previously denoted, this data set is based on a stochastic view of the popularity concept, and therefore the data concerns the aggregate behaviour of users. This data is obtained from public end-points of the social media sources, which is already anonymized and aggregated by the data proprietaries. Thus, this data set does not include any personally identifiable information either from single or groups of users, e.g. social network data, guaranteeing the privacy of the users.


\section{User Notes}\label{sec:usernotes}

In order to facilitate the quick use of the data set a file with basic operations was designed in the programming language \textbf{R}. This file is located in the data set archive.


\section*{Acknowledgments}\label{sec:ack}

This work is financed by the ERDF – European Regional Development Fund through the COMPETE 2020 Programme within project POCI-01-0145-FEDER-006961, and by National Funds through the FCT – Fundação para a Ciência e a Tecnologia (Portuguese Foundation for Science and Technology) as part of project UID/EEA/50014/2013.


\bibliographystyle{plainnat}
\bibliography{sample}

\end{document}